\documentclass[12pt]{article}
\usepackage{feynmp,epsfig,a4}
\newcommand{\lsim}{\raisebox{-4pt}{$\,\stackrel{\textstyle
                                                         <}{\sim}\,$}}

\newcommand{\nn}{\nonumber}
\newcommand{\gev}{{\rm GeV}}
\renewcommand{\d}{\rm{d}}
\newcommand{\als}{\alpha_{\rm s}}
\newcommand{\aem}{\alpha_{\rm em}}
\newcommand{\vk}{{\bf k}_\perp}
\newcommand{\vd}{{\bf \Delta}_\perp}
\newcommand{\wf}{wave function}
\newcommand{\ibid}[1]{{\it ibid.}~#1}

\begin{document}
\begin{fmffile}{comppic}

\begin{flushright}
DESY 99-025 \\
WUB 99-5 \\
FNT/T-99/04 \\
hep-ph/9903268 \\
\end{flushright}

\begin{center}
\vskip 3.5\baselineskip
\textbf{\Large Skewed Parton Distributions in \\[0.3\baselineskip]
  Real and Virtual Compton Scattering}
\vskip 2.5\baselineskip
M.~Diehl$^{1}$, Th.~Feldmann$^{2}$, R.~Jakob$^{3}$ and P.~Kroll$^{2}$
\vskip \baselineskip
1. Deutsches Elektronen-Synchroton DESY, D-22603 Hamburg, Germany \\
2. Fachbereich Physik, Universit\"at Wuppertal, D-42097 Wuppertal,
   Germany \\ 
3. Universit\`{a} di Pavia and INFN, Sezione di Pavia, 
   I-27100 Pavia, Italy
\vskip 3\baselineskip
\textbf{Abstract} \\[0.5\baselineskip]
\parbox{0.9\textwidth}{The handbag contribution to Compton scattering
  at moderately large momentum transfer factorises into parton-photon
  subprocess amplitudes and new form factors representing
  $1/x$-moments of skewed parton distributions. A detailed
  phenomenological study for polarised and unpolarised real and
  virtual Compton scattering is presented.}
\vskip 1.5\baselineskip
\end{center}


\noindent
{\bf 1.} The interest in the interplay between hard inclusive and
exclusive reactions has recently been revived by theoretical work on
deeply virtual Compton scattering (DVCS) and skewed parton
distributions (SPDs) \cite{mue98}. These SPDs are hybrid objects,
which combine properties of form factors and of ordinary parton
distributions. In fact, reduction formulas reveal the close connection
of these quantities. It has also been shown recently
\cite{rad98a,DFJK} that at moderately large momentum transfer real and
virtual Compton scattering off protons approximately factorises into a
hard parton-photon subprocess and a soft proton matrix element
described by new form factors specific to Compton scattering. These
new form factors, as the ordinary electromagnetic ones, represent
moments of SPDs and can be modelled by overlaps of light-cone \wf{}s
\cite{rad98a,DFJK}, which provide the link between exclusive and
inclusive reactions. In this overlap representation, which implies
Feynman's end-point mechanism, the SPDs are given as products of
ordinary parton distributions and exponentials of $t\, (1-x)/x$,
provided one makes a simplifying assumption about the \wf{}s. Here $t$
is the squared momentum transfer experienced by the proton, and $x$
the usual fraction of the light-cone plus component of the proton
momentum carried by the active parton, i.e.\ the one entering the
parton-photon subprocess.

It is to be emphasised that the soft physics mechanism is
complementary to the perturbative one \cite{bro80}, and that both
contributions have to be taken into account. We argue, however, that
for large angle Compton scattering the soft contribution, although
formally representing a power correction to the asymptotically leading
perturbative one,\footnote{In this respect factorisation in large
  angle Compton scattering is not on the same footing as the one in
  DVCS, where the factorising diagrams are dominant for asymptotically
  large photon virtuality, and where factorisation can be proven to
  all orders in perturbation theory.} dominates at experimentally
accessible momentum transfers. For electromagnetic nucleon form
factors it has been shown that agreement with the data can be achieved
by calculating both hard scattering and soft overlap contributions
with a moderately asymmetric wave function, and the soft contribution
was indeed found to dominate for $-t$ of order $10~\gev^2$
\cite{bol96}. The soft contribution to large angle Compton scattering,
evaluated with the same \wf, is also in reasonable agreement with
available data \cite{DFJK}. The perturbative contribution has been
calculated in \cite{van97,kron91} to leading twist accuracy and is way
below the Compton data unless strongly asymmetric, i.e.\ end-point
concentrated distribution amplitudes are used. These give however
results dominated by contributions from the soft end-point regions,
where the assumptions of a leading twist perturbative calculation
break down, and have also been criticised on other grounds, cf.\ for
instance \cite{bol96,isg}. {}From the results of \cite{bol96,van97} we
estimate that the perturbative contribution to Compton scattering
amounts to less than $10\%$ of the data for $-t$ in the region of a
few $10~\gev^2$.

The data of many exclusive observables exhibit approximate dimensional
counting rule behaviour, a fact that is frequently considered as
evidence for the dominance of perturbative physics. In our opinion,
this conclusion is unjustified: The running of $\als$ and the
evolution of the hadronic \wf{}s often provide large powers of
$\ln{s}$, which should modify the dimensional counting rule behaviour
substantially. Such modifications are however not seen in the data.
In these reactions the effective scale of hardness is typically rather
low, so that the effect of the logarithms should be especially strong.
One may argue that the effective scale in these cases is so small that
the running coupling becomes frozen.  This indicates, however, that
one is not in the perturbative regime (the freezing of $\als$ being
certainly a nonperturbative effect), and also means that power
corrections can be large. In the soft physics approach, on the other
hand, approximate dimensional counting rule behaviour holds in a
limited range of momentum transfer, which is controlled by the
transverse size of the hadrons involved. For electromagnetic form
factors and Compton scattering this mimicked scaling behaviour is well
in agreement with experiment \cite{rad98a,DFJK,bol96}. Naturally the
question arises how to interpret the approximate dimensional counting
rule behaviour in other exclusive reactions, such as proton-proton
elastic scattering. A tentative answer to this question will be given
in this article.

The main purpose of this paper is however to present a
phenomenological study of real and virtual Compton scattering in the
soft physics approach in order to facilitate comparison with other
theoretical results on this reaction and with future experimental data
that might be obtained at Jefferson Lab or at an ELFE-type accelerator
at DESY or CERN.

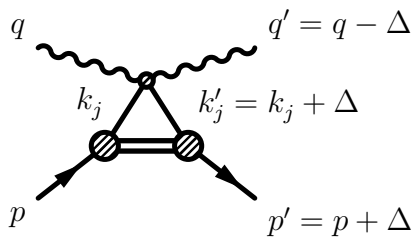
\begin{figure}[hbtp]
\parbox{\textwidth}{\begin{center}
\unitlength0.8cm
\fmfframe(0.5,1.0)(3.0,1.0){
\begin{fmfgraph*}(4.5,2.5)
\fmfpen{thick}
\fmfleft{Q1,dummy1,p1}\fmfright{Q2,dummy3,p2}
\fmf{fermion,tension=1.25}{Q1,v1} \fmf{double,tension=0.5}{v1,v2}
\fmf{fermion,tension=1.25}{v2,Q2} 
\fmf{plain}{v1,v3}
\fmf{plain}{v3,v2}
\fmfv{label=$k^{\phantom{}}_j~$,la.a=111,la.di=0.10w}{v1}
\fmfv{label=$k_j'=k^{\phantom{}}_j+\Delta$,la.a=67,la.di=0.09w}{v2}
\fmf{photon,tension=2}{p1,v3} \fmf{photon,tension=2}{p2,v3}
\fmfv{label=$p$}{Q1}\fmfv{label=$p'=p+\Delta$}{Q2}
\fmfv{label=$q$}{p1}\fmfv{label=$q'=q-\Delta$}{p2}
\fmfblob{.1w}{v1}\fmfblob{.1w}{v2}\fmfblob{.05w}{v3}
\end{fmfgraph*}}
\end{center}}
\vspace{-0.8cm}
\caption{\label{fig1} The handbag diagram for Compton scattering off
  protons. The horizontal lines represent any number of spectator
  partons. }
\end{figure}

\vskip\baselineskip
\noindent
{\bf 2.} Let us briefly outline the calculation of large angle Compton
scattering in the soft physics approach; for details we refer to
\cite{DFJK}. The amplitude is evaluated from the handbag diagram shown
in Fig.\ \ref{fig1}. The large blobs denote soft proton \wf{}s, i.e.\ 
\wf{}s with their perturbative tails removed, and the small blob
attached to the photon lines represents the elementary subprocess,
Compton scattering off quarks or antiquarks, which is calculated in
lowest order QED with point-like quark-photon couplings.  The physical
situation is that of a hard photon-parton scattering and the soft
emission and reabsorption of a parton by the hadron, as in the
familiar handbag diagram for DVCS or inclusive deeply inelastic
scattering~(DIS).

Four-momenta are also defined in Fig.\ \ref{fig1}. As usual we define
the Mandelstam variables $s=(p+q)^2$, $t=\Delta^2$, $u=(p-q')^2$, and
write $Q^2=-q^2$ for the incoming photon virtuality and $m$ for the
proton mass. We require $s$, $-u$ and $-t$ to be large on a hadronic
scale, which defines large angle scattering and the region of validity
of our calculation.

We assume that soft hadron wave functions are dominated by parton
virtualities in the range $|k_i^2| \lsim \Lambda^2$, where $\Lambda$
is a hadronic scale in the GeV region, and by intrinsic transverse
parton momenta $\vk{}_i$ (defined with the respect to their parent
hadron's momentum) that satisfy ${\bf k}_{\perp i}^2
/x_i^{\phantom{.}} \lsim \Lambda^2$. This leads to an approximate
equality of the Mandelstam variables in the parton-photon subprocess
and the overall proton-photon reaction up to corrections of order
$\Lambda^2\, (1-Q^2/t)$. Therefore the parton-photon scattering is
hard, and when calculating it we approximate the momenta
$k_j^{\phantom{.}}$, $k'_j$ of the active partons as being on shell,
collinear with their parent hadrons and with light cone fractions
$x_j^{\phantom{.}} = x'_j = 1$.

Under the above assumptions the helicity amplitudes ${\cal
  M}_{\mu'\nu',\,\mu \nu}$ for large angle Compton scattering can be
written in terms of soft proton matrix elements and hard parton-photon
scattering amplitudes ${\cal H}_{\mu'\lambda',\,\mu\lambda}$. We
define proton and photon helicities in the photon-proton c.m., which
is convenient for phenomenological applications and comparison with
other results. The amplitudes conserving the proton helicity are
explicitly given by
\begin{equation}
{\cal M}_{\mu'+,\,\mu +} = \;2\pi\aem \left[{\cal
    H}_{\mu'+,\,\mu+}\,(R_V + R_A)\,
  + \, {\cal H}_{\mu'-,\,\mu-}\,(R_V - R_A) \right ]\,,
\label{final}
\end{equation}
proton helicity flip will be discussed shortly. $\mu$ and $\mu'$ are
the helicities of the incoming and outgoing photon, and $\nu,\,
\lambda$ and $\nu',\, \lambda'$ those of the incoming and outgoing
proton and parton, respectively. From parity invariance one has ${\cal
  M}_{\mu'\nu,\,\mu\nu} = (-1)^{\mu'-\mu} {\cal
  M}_{-\mu'-\nu,\,-\mu-\nu}$ and an analogous equation for ${\cal
  H}_{\mu'\lambda,\,\mu\lambda}$. For the sake of legibility we label
explicit helicities only by their signs, i.e.\ in the matrix elements
we write $+$, $-$ instead of $+1/2$, $-1/2$ for fermions.

Since the partons are taken as massless there is no parton helicity
flip in the photon-parton subprocess amplitudes, which read 
\begin{eqnarray} 
{\cal H}_{++,\,++} &=&   2\, \sqrt{\frac{s}{-u}}\: \frac{s+Q^2}{s}
\,, \hspace{3em}
{\cal H}_{-+,\,-+} \:=\: 2\, \sqrt{\frac{-u}{s}}\: \frac{s}{s+Q^2}
\,, \nonumber \\
{\cal H}_{-+,\,++} &=&   2\, \frac{Q^2}{s+Q^2}\: \frac{t}{\sqrt{-s u}}
\,, \hspace{2.7em}
{\cal H}_{-+,\,0+} \:=\: - 2\: \frac{Q}{s+Q^2}\: \sqrt{-2t}\,,
\label{hard-amplitudes}
\end{eqnarray}
and ${\cal H}_{++,\,-+} \,=\,{\cal H}_{++,\,0+} \,=\, 0.$ Within the
accuracy of our calculation the subprocess amplitudes are real;
$\als$-corrections, however, will lead to non-zero imaginary parts.

The soft proton matrix elements in Eq.\ (\ref{final}), $R_V$ and
$R_A$, represent form factors specific to Compton scattering
\cite{rad98a,DFJK}. $R_V$ is defined by
\begin{eqnarray} 
\lefteqn{ \sum_a e_a^2\, \int_0^1\, \frac{{\d} x}{x}\, p^+
   \int {{\d} z^-\over 2\pi}\, e^{i\, x p^+ z^-}
     \langle p',\nu'|\,
     \overline\psi{}_{a}(0)\, \gamma^+\,\psi_{a}(z^-) - 
     \overline\psi{}_{a}(z^-)\, \gamma^+\,\psi_{a}(0) 
     \,| p,\nu\rangle } \nonumber \\
&& \hspace{8em} = R_V(t)\, \bar{u}(p',\nu')\, \gamma^+ u(p,\nu)\,
                   + R_T(t)\, \frac{i}{2m}\bar{u}(p',\nu')
                          \sigma^{+\rho}\Delta_\rho u(p,\nu)\,,
\label{R-form-factors}
\end{eqnarray}
where the sum runs over quark flavours $a$ ($u$, $d$, \ldots), $e_a$
being the electric charge of quark $a$ in units of the positron
charge. The matrix element is to be evaluated in a frame where the
light-cone plus momentum of the proton is unchanged, i.e.\ where
$\Delta^+=0$. There is an analogous equation for the axial vector
proton matrix element, which defines the form factor $R_A$. Note that,
as in DIS and DVCS, only the plus components of the proton matrix
elements enter in the Compton amplitude, which is a nontrivial
dynamical feature given that, in contrast to DIS and DVCS, not only
the plus components of the proton momenta but also their minus and
transverse components are large now. Due to time reversal invariance
the form factors $R_V$, $R_A$ etc.\ are real functions.  As the
definition (\ref{R-form-factors}) reveals they are $1/x$ moments of
SPDs at zero skewedness parameter $\zeta=-\Delta^+/p^+$ \cite{mue98}.

Some remarks on the proton spin are in order. The description of $R_T$
or of its electromagnetic counterpart $F_2$ involves components in the
proton wave function, where the parton helicities do not add up to the
helicity of the hadron, whose modelisation is beyond the scope of this
work. It is however natural to assume that $R_T /R_V \sim F_2 /F_1$,
and the latter ratio is known to be $F_2/F_1 \simeq -m^2/t$ at large
$t$. Terms going with $R_T$ in proton helicity non-flip amplitudes are
then corrections of order $m^2/t$ and have been omitted in Eq.\ 
(\ref{final}), given that already our evaluation of the handbag
diagrams is only accurate up to corrections in $\Lambda^2/t$.  For
proton helicity flip we obtain amplitudes going with $R_T
{\sqrt{-t/m^2}}$ and $R_{V,A} {\sqrt{-m^2/t}}$, which are down
compared with non-flip amplitudes by a factor of ${\sqrt{-m^2/t}}$.
Within our accuracy observables involving unpolarised and
longitudinally polarised protons can thus be calculated from the
proton helicity non-flip amplitudes (\ref{final}) alone.

\vskip\baselineskip
\noindent
{\bf 3.} Before we present numerical results for the observables of
real and virtual Compton scattering we have to model the new form
factors. In a frame where $\Delta^+=0$ the form factors $R_V$ and
$R_A$ can be represented by overlaps of light-cone \wf{}s summed over
all Fock states, in close analogy with the famous Drell-Yan formula
\cite{DY} for the electromagnetic form factor:
\begin{equation}
R_{V,A}(t) = \sum_{N,\beta,j} \,
      \kappa_{V,A}^{\phantom{.}}\, e^2_j\,
      \int\, [{\d} x]_N [{\d}^2 \vk]_N\, \frac{1}{x_j}\,
      \Psi^*_{N\beta}(x_i^{\phantom{.}},{\bf k}'_{\perp i})\,
      \Psi_{N\beta}(x_i,\vk{}_i)
\label{DYgen}
\end{equation}
with $\kappa_V=1$ and $\kappa_A=2\lambda_j$. Each Fock state $N$ is
described by a number of terms, each with its own momentum space \wf\ 
$\Psi_{N\beta}$, where $\beta$ labels different spin-flavour
combinations of the partons. The sum over the active parton, $j$, with
charge $e_j$ and helicity $\lambda_j$ runs over all partons in a given
Fock state. Primed and unprimed
intrinsic transverse momenta are related to each other by ${\bf
  k}'_{\perp i} = \vk{}_i \,-\, x_i\, \vd$ for $i\neq j$ and ${\bf
  k}'_{\perp j} = \vk{}_{j}\,+\, (1-x_j)\, \vd$, and $[{\rm d}x]_N
[{\rm d}^2 \vk]_N$ is the $N$-particle integration measure, cf.\ 
\cite{DFJK}. Assuming a simple Gaussian $\vk{}_i$-dependence of the
soft Fock state \wf{}s,
\begin{equation}
   \Psi_{N\beta}(x_i,\vk{}_i) \propto  \exp
     \left [
      -a_N^{2} \sum_{i=1}^{N} \frac{{\bf k}_{\perp i}^2}{x_{i}}
     \right ]\,,
\label{BLHMOmega}
\end{equation}
one can explicitly carry out the momentum integrations in
(\ref{DYgen}). The ansatz (\ref{BLHMOmega}) satisfies various
theoretical requirements \cite{chi95,bro98} and is in line with our
hypothesis that the soft hadronic \wf{}s are dominated by transverse
momenta with ${\bf k}_{\perp i}^2 /x_i^{\phantom{.}} \lsim \Lambda^2$,
necessary to achieve the factorisation of the Compton amplitudes into
soft and hard parts. The results of the transverse momentum
integrations for $R_V$ and $R_A$ are respectively related with the
Fock state contributions to the unpolarised and polarised parton
distribution functions. For simplicity one may further assume a common
transverse size parameter $a_N=\hat a$ for all Fock
states.\footnote{Note that we restrict ourselves to large values of
  $t$ here, where the main contribution to the overlap integral
  (\protect\ref{DYgen}) is only due to a limited number of Fock
  states.} This immediately allows one to sum over them, without
specifying the $x_i$-dependence of the \wf{}s. One then arrives at
\begin{equation}
R_V(t) = \sum_a\, e_a^2\, \int \frac {{\d} x}{x}\, 
         \exp{\left[\frac12 \hat a^2 t \frac{1-x}{x}\right]}\; 
         \{ q_a(x) + \bar{q}_a(x) \} \,,
\label{ffspd}
\end{equation}
and the analogue for $R_A$ with $q_a+\bar{q}_a$ replaced by $\Delta
q_a + \Delta \bar{q}_a$. The result (\ref{ffspd}) is very instructive
as it elucidates the link between the parton distributions of DIS and
exclusive reactions. Evaluating the form factors from the parton
distributions of GRV \cite{GRV} and with $\hat a = 1~\gev^{-1}$, one
already finds results for the Compton cross section in fair agreement
with experiment. In order to improve on the approximation
(\ref{ffspd}) the lowest three Fock states were modelled explicitly in
\cite{DFJK}, assuming specific distributions amplitudes and fitting
the \wf{} parameters to the GRV parton distributions at $x>0.5$. In
the present letter we make use of the model form factors as given
there.

These form factors behave as $1/t^2$ in the momentum transfer range
from about 5 to 15~\gev$^2$ and, consequently, the Compton cross
section shows approximate $s^{-6}$ scaling behaviour for photon
energies in the region of several GeV, cf.\ \cite{DFJK}. With
increasing $-t$ the form factors gradually turn into the soft physics
asymptotics $\propto 1/t^4$, which follows from the $x_i$-dependence
of the model wave functions at the end points. In that region of $t$
the perturbative contribution will take the lead.

\vskip\baselineskip
\noindent
{\bf 4.} The contribution of virtual Compton scattering to the
unpolarised  $ep\to ep\gamma$ cross section can be decomposed into
four partial cross sections (for details see \cite{kro96}): the cross
sections for transverse photons (reducing to the unpolarised cross
section for real Compton scattering, i.e.\ for  $Q^2=0$) and for
longitudinal photons,
\begin{eqnarray}
\frac{{\d} \sigma_{\rm T}}{{\d} t} &=& \frac{1}{32\pi s (s+Q^2)}\; 
       \sum_{\mu',\nu',\nu}\; |{\cal M}_{\mu'\nu',\,+\nu}|^2\,, \nn\\
\frac{{\d} \sigma_{\rm L}}{{\d} t} &=& \frac{1}{32\pi s (s+Q^2)} \;
       \sum_{\mu',\nu',\nu}\; |{\cal M}_{\mu'\nu',\,0\,\nu} |^2\,, 
\label{tandl}
\end{eqnarray}
and the transverse-transverse and longitudinal-transverse interference
terms
\begin{eqnarray}
\frac{{\d} \sigma_{\rm TT}}{{\d} t} &=& -\frac{1}{64\pi s (s+Q^2)}\; 
            {\rm Re}\,\sum_{\mu',\nu',\nu}\; 
            {\cal M}_{\mu'\nu',\,+\nu}^\ast\, 
            {\cal M}_{\mu'\nu',\,-\nu}^{\phantom{\ast}} \,, \nn\\
\frac{{\d} \sigma_{\rm LT}}{{\d} t} &=& 
    -\frac{\sqrt{2}}{64\pi s (s+Q^2)}\; 
            {\rm Re}\, \sum_{\mu',\nu',\nu}\; 
            {\cal M}_{\mu'\nu',\,0\,\nu}^\ast\,
                 [{\cal M}_{\mu'\nu',\,+\nu}^{\phantom{\ast}}- 
                  {\cal M}_{\mu'\nu',\,-\nu}^{\phantom{\ast}}]\,.
\label{inter}
\end{eqnarray}
In certain kinematical regions, namely for small values of $-t/Q^2$ or
of the ratio $\varepsilon$ of longitudinal and transverse photon flux
in the Compton process, the full $ep\to ep\gamma$ cross section
receives substantial contributions from the Bethe-Heitler process,
where the final state photon is radiated by the electron. This process
is completely calculable for values of $t$ where the elastic proton
form factors $F_1$ and $F_2$ are known, and in suitable kinematics its
interference with the Compton process can be used to study the latter
at amplitude level.

For real Compton scattering a number of polarisation observables have
been introduced \cite{rol79}. Of particular interest is the initial
state helicity correlation
\begin{equation}
A_{\rm LL}\, \frac{{\d} \sigma}{{\d} t} \:=\:
    \frac12\,\left[\frac{{\d} \sigma(++)}{{\d} t}-\frac{{\d}
    \sigma(+-)}{{\d} t}\right] 
    \:=\:  \frac{1}{32\pi s^2}\, \sum_{\mu',\nu'}\;
           \left[|{\cal M}_{\mu'\nu',\,++}|^2 -
                 |{\cal M}_{\mu'\nu',\,+-}|^2 \right] \,,
\label{all}
\end{equation}
which , when $R_T$ and mass corrections are neglected, measures the
product $R_V R_A$, while the unpolarised  cross section receives
contributions from $R_V^2$ and $R_A^2$  only:
\begin{eqnarray}
\frac{{\d} \sigma}{{\d} t} &=& \frac{2\pi\aem^2}{s^2} \, 
 \left[\, \frac{1}{2} (R_V^2+R_A^2) 
          \left(-\frac{s}{u} - \frac{u}{s}\right)
        + (R_V^2-R_A^2) \,\right] \,, \nn \\
A_{\rm LL}\, \frac{{\d} \sigma}{{\d} t} &=& \frac{2\pi\aem^2}{s^2} \;
  R_V R_A \left(\frac{u}{s} - \frac{s}{u}\right) \,.
\end{eqnarray}

Other spin observables for real Compton scattering are the incoming
photon asymmetry $\Sigma$ and the helicity transfer parameter 
$D_{\rm LL}$,
\begin{eqnarray}
\Sigma\,\frac{{\d} \sigma}{{\d} t} \:=\:
   \frac12\,\left[\frac{{\d} \sigma_\perp}{{\d} t}-
                  \frac{{\d} \sigma_\parallel}{{\d} t}\right] \:=\:
   \frac{1}{32\pi s^2}\; {\rm Re}\, \sum_{\mu',\nu',\nu}\;
   {\cal M}_{\mu'\nu',\,+\nu}^\ast\, 
   {\cal M}_{\mu'\nu',\,-\nu}^{\phantom{\ast}}\,,
                             \nn\\
D_{\rm LL}\, \frac{{\d} \sigma}{{\d} t} \:=\:
             \frac{{\d} \sigma(+,+)}{{\d} t}-
             \frac{{\d} \sigma(+,-)}{{\d} t} \:=\:
             \frac{1}{32\pi s^2}\, \sum_{\nu',\nu}\;
          \left[|{\cal M}_{+\nu',\,+\nu}|^2 -
                |{\cal M}_{+\nu',\,-\nu}|^2 \right] \,
\label{sigma-dll}
\end{eqnarray}
where $\perp$ and $\parallel$ respectively refer to linear photon
polarisation normal to and in the scattering plane. For real Compton
scattering the photon helicity turns out to be strictly conserved in
the soft physics approach (up to possible $\als$ corrections), cf.\ 
Eq.\ (\ref{hard-amplitudes}), so that $\Sigma$ and $D_{\rm LL}$
acquire the values 0 and $1$, respectively. In the diquark model, a
variant of the standard perturbative Brodsky-Lepage approach to
exclusive reactions \cite{bro80}, small deviations from these values
have been obtained \cite{kro91} since the photon helicity flip
amplitudes are non-zero, although suppressed by powers of $1/s$ in
this model. The leading twist hard scattering results of \cite{kron91}
deviate only slightly from our soft physics values.
 
\begin{figure}[hbtp]
\begin{center}
\psfig{file=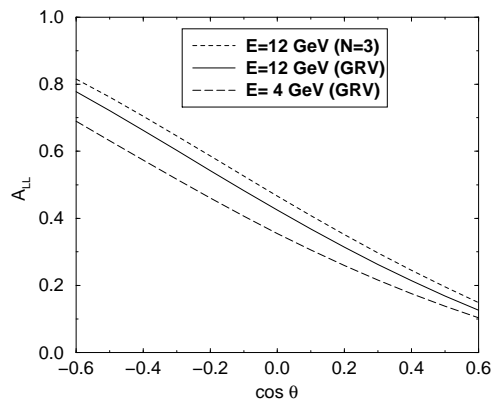, bb=95 45 565 625, width =5.2cm, angle=-90}   
\end{center}
\caption{\label{fig:all} Predictions for the initial state helicity 
  correlation $A_{\rm LL}$ at photon energies of 4~\gev{}
  (long-dashed) and 12~\gev{} (solid line) in the proton rest frame
  vs.\ $\cos{\theta}$, where $\theta$ is the c.m.\ scattering angle.
  The short-dashed line represents the contribution from the valence
  Fock state alone at 12~\gev. $A_{\rm LL}$ is evaluated with the
  Compton form factors calculated in \protect\cite{DFJK}.}
\end{figure}

Making use of the numerical results for $R_V$ and $R_A$ given in
\cite{DFJK} (cf.\ Sect.\ 3) we evaluate the initial state helicity
correlation $A_{\rm LL}$ (\ref{all}) for real Compton scattering. Its
$\cos{\theta}$\/-dependence (cf.\ Fig.\ \ref{fig:all}) is roughly
given by $(s^2- u^2)/(s^2+u^2)$, and reflects that of the
corresponding helicity correlation for the photon-parton subprocess.
It is opposite in sign to the diquark model predictions \cite{kro91}.
In the leading twist hard scattering approach the two available
results \cite{van97,kron91} strongly differ from each other and from
the result presented here.

\begin{figure}[hbtp]
\begin{center}
\psfig{file=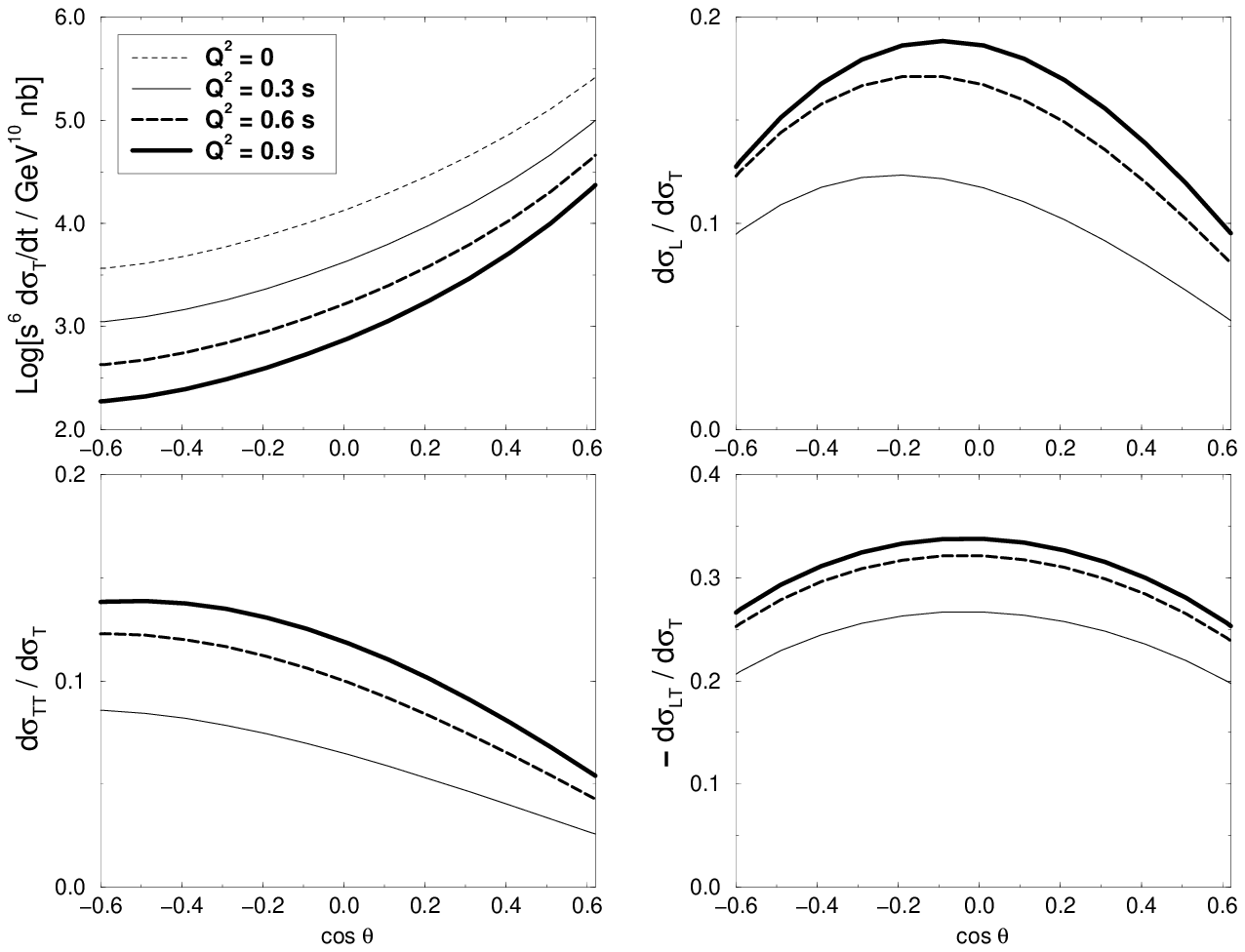, bb= 115 445 480 725,  width =10.8cm}   
\end{center}
\vspace{-0.8cm}
\caption{\label{fig:vir} Predictions for the virtual Compton cross
  sections at $s=10$~GeV$^2$ and different values of $Q^2$  vs.\ 
  $\cos{\theta}$.}
\end{figure}

Predictions for the various cross sections for virtual Compton
scattering are shown in Fig.~\ref{fig:vir} at a photon energy of
5~\gev{} in the proton rest frame and for a set of $Q^2$ values.
Comparing with the only other available results, namely those from the
diquark model \cite{kro96}, we see that the transverse cross section
in both approaches comes out rather similar, while the other three
cross sections are generally larger and with a smoother
$Q^2$-dependence in the soft physics approach than in the diquark
model. In contrast to the diquark model the transverse-transverse
interference term is strictly zero now in the limit $Q^2=0$, where the
ratio ${\d}\sigma_{\rm TT}/{\d}\sigma_{\rm T}$ is equivalent to the
photon asymmetry $\Sigma$ defined in Eq.\ (\ref{sigma-dll}). We note
that in the soft physics approach ${\d}\sigma_{\rm LT}
/{\d}\sigma_{\rm T} \le 0$ as long as $R_V^2 \ge R_A^2$, which is
satisfied with our representation (\ref{ffspd}) in terms of parton
distributions.

\begin{figure}[hbtp]
\begin{center}
\psfig{file=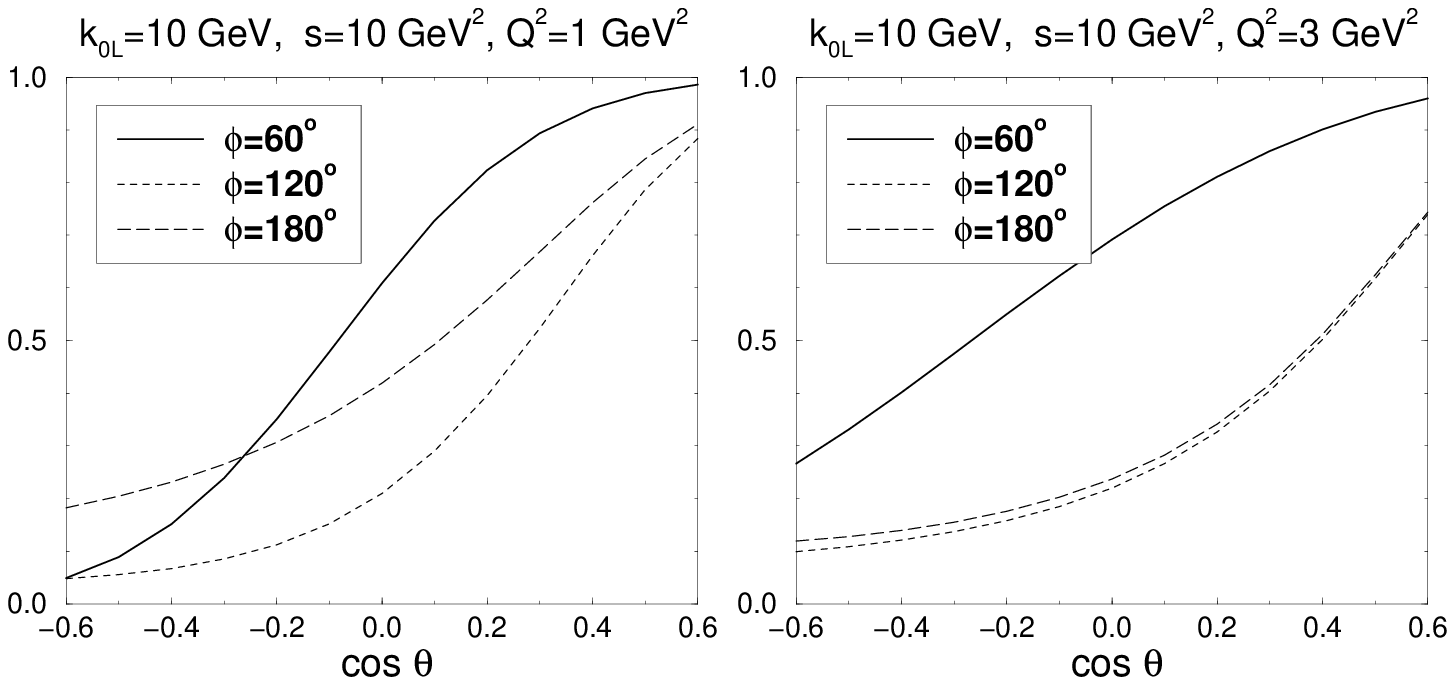, bb= 90 525 515 725,  width =10.8cm}   
\end{center}
\vspace{-0.8cm}
\caption{\label{fig:bh} The difference between the full $ep\to
  ep\gamma$ cross section and the contribution of the Compton process
  alone, divided by the full cross section, as a function of
  $\cos{\theta}$ at fixed $k_{0L}$, $s$, $Q^2$ and $\phi$. The
  azimuthal angle $\phi$ between the electron and hadron planes is
  defined in \protect\cite{kro96}, and $k_{0L}$ denotes the electron
  beam energy in the proton rest frame.}
\end{figure}

Fig.\ \ref{fig:bh} shows the difference between the full $ep\to
ep\gamma$ cross section and the contribution of the Compton process
alone, divided by the full cross section. As expected the
Bethe-Heitler process becomes dominant for increasing $\cos{\theta}$,
i.e.\ for decreasing $-t$. For the kinematics considered here the
relative importance of Bethe-Heitler and Compton is similar to the one
found in the diquark model \cite{kro96}. A detailed investigation of
the interplay between the Bethe-Heitler and Compton contributions and
their interference as a function of the various kinematical variables
is beyond the scope of this letter.

In Ref.\ \cite{kro96} the relevance of the beam asymmetry for $ep\to
ep\gamma$
\begin{equation}
A_{\rm L} = \frac{ {\d}\sigma (+) - {\d}\sigma (-)} 
                 { {\d}\sigma (+) + {\d}\sigma (-)} \,,
\end{equation}
where the labels $+$ and $-$ denote the lepton beam helicity, has been
pointed out.  It is sensitive to the imaginary part of the
longitudinal-transverse interference in the Compton process, while
${\d}\sigma_{\rm LT}/{\d}t$ measures its real part. In the diquark
model \cite{kro96} the virtual Compton contribution to $A_{\rm L}$ is
very small but the full asymmetry is spectacularly enhanced in regions
of strong interference between the Compton and the Bethe-Heitler
amplitudes. In these regions $A_{\rm L}$ essentially measures the
relative phase between the complex virtual Compton amplitudes and the
real Bethe-Heitler ones. In the standard perturbative approach a
non-zero value of $A_{\rm L}$ is also to be expected in the
interference region because of the perturbatively generated phases of
the Compton amplitudes. In the soft physics approach, on the other
hand, $A_{\rm L}$ is zero since all amplitudes are real within the
accuracy of our calculation. Due to $\als$-corrections in the
photon-parton subprocess $A_{\rm L}$ may become non-zero in the soft
physics approach.\footnote{The cat's ears diagrams with a hard gluon
  (cf.\ \protect\cite{DFJK}) can also give imaginary parts to the
  Compton amplitudes.}

One may finally consider the transverse polarisation (normal to the
scattering plane) of the initial or the final state proton. As is well
known the corresponding polarisation asymmetry requires both
non-vanishing proton helicity flip amplitudes and relative phases
between flip and non-flip amplitudes.  In the soft physics mechanism
all amplitudes are approximately real. Moreover, as we argued above,
the helicity flip amplitudes are suppressed compared to the non-flip
ones by a factor ${\sqrt{-m^2/t}}$. We therefore predict very small
proton polarisations in the soft physics approach. Because of hadron
helicity conservation the transverse proton spin asymmetries are zero
in the standard perturbative approach \cite{kron91}. Only the diquark
model provides both necessary ingredients and predicts proton
polarisations of up to $10\%$ \cite{kro91}.

\vskip\baselineskip
\noindent
{\bf 5.} To summarise, the detailed predictions for real and virtual
Compton scattering obtained from the handbag diagram exhibit
interesting features and characteristic helicity dependences.
Comparison with perturbative calculations, either obtained within the
standard hard scattering approach or its diquark variant, reveals
marked differences which may allow one to distinguish between these
mechanisms experimentally. Data for these observables from Jefferson
Lab or other accelerators are eagerly awaited.

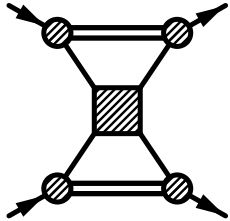
\begin{figure}[hbtp]
\begin{center}
\unitlength0.8cm
\fmfframe(0.1,0.1)(0.1,0.1){
\begin{fmfgraph}(4.5,3.5)
\fmfpen{thick}
\fmfleft{Q1,Q2}\fmfright{P1,P2}
\fmf{fermion,tension=2.0}{Q1,v1} 
  \fmf{double,tension=0.5}{v1,v2}
  \fmf{fermion,tension=2.0}{v2,P1} 
\fmf{fermion,tension=2.0}{Q2,v3} 
  \fmf{double,tension=0.5}{v3,v4}
  \fmf{fermion,tension=2.0}{v4,P2} 
\fmfpolyn{shaded}{i}{4}
\fmf{plain}{v1,i1}
\fmf{plain}{i2,v2} 
\fmf{plain}{v3,i4}
\fmf{plain}{i3,v4} 
\fmfblob{.1w}{v1}\fmfblob{.1w}{v2}
\fmfblob{.1w}{v3}\fmfblob{.1w}{v4}
\end{fmfgraph}}
\end{center}
\caption{\label{fig:pp} The soft physics mechanism for $pp$ elastic 
  scattering. The horizontal lines represent any number
  of spectator partons.}
\end{figure}

It is particularly interesting that the soft physics approach can
account for the experimentally observed approximate dimensional
counting rule behaviour, at least for Compton scattering and for form
factors. This tells us that it is premature to infer the dominance of
perturbative physics from the observed scaling behaviour . One may
object that the perturbative explanation (leaving aside the logarithms
from the running of $\als$ and from the evolution) works for many
exclusive reactions, while in the soft physics approach the
approximate counting rule behaviour is accidental, depending on
specific properties of a given reaction. In our opinion, and we are
going to substantiate this briefly, the approximate counting rule
behaviour is an unavoidable feature of the soft physics approach. Let
us consider proton-proton elastic scattering at $s, \; -t,\; -u \gg
\Lambda^2$. Viewing this process as in Fig.\ \ref{fig:pp} we recognise
the factorisation into soft hadron matrix elements and a hard
scattering of spin 1/2 partons.\footnote {Contributions where the
  active partons are gluons are suppressed since the soft hadron
  \wf{}s provide less gluons with large momentum fraction $x$, cf.\ 
  \protect\cite{DFJK}.} This model for hadron-hadron scattering bears
resemblance to the parton scattering model invented long time ago
\cite{arb69}. The hard scattering kernels are dimensionless and
therefore depend only on the ratio $t/s$ and on the parton helicities.
The soft hadron matrix elements represent form factors similar to the
electromagnetic or the Compton form factors, except that the parton
charges do not appear.

All these form factors are smooth functions of the momentum transfer
and, when scaled by $t^2$, exhibit a broad maximum in the $-t$-range
from about 5 to 15~\gev$^2$, set by the transverse hadron size, i.e.\ 
by a scale of order 1~\gev$^{-1}$. To see how the maximum can be at
$-t$ quite above a GeV$^2$ consider for example the form factor
$R_V(t)$ as given in (\ref{ffspd}). For the position of the maximum of
$t^2 R_V(t)$ we obtain
\begin{equation}
- t = 4 \hat a^{-2}\, \left\langle { \frac{1-x}{x}} \right\rangle^{-1}
\label{maxpos}
\end{equation}
where the $t$-dependent mean value $\langle \frac{1-x}{x} \rangle$ is
defined by weighting with the integrand of (\ref{ffspd}). It comes out
around $0.5$ at the value of $t$ where the maximum is taken. Note also
that both sides of the implicit equation (\ref{maxpos}) increase with
$-t$. It is thus approximately satisfied over a certain $t$-range, in
other words the maximum of the scaled form factor is quite broad.

Without a full-fledged analysis, i.e.\ without specifying the hard
scattering kernels ${\cal H}(t/s)$, it is clear now that the soft
physics approach provides approximate $s^{-10}$ scaling of ${\d}\sigma
/{\d}t$ in proton-proton scattering at large, fixed scattering angle
and $s$ in the range from $10~\gev^2$ to $30~\gev^2$.  The agreement
of this prediction with experiment \cite{pp} is reasonable as we
checked. We remind however the reader that the proton-proton data show
fluctuations superimposed to the $s^{-10}$ behaviour. These
fluctuations, if a real dynamical feature, tell us that there still is
another momentum scale relevant in that kinematical region,
contradicting the very idea of dimensional scaling.

For mesons all our arguments apply in a similar fashion. The
corresponding form factors approximately behave like $1/t$ over a
certain range of $t$, again mimicking dimensional counting. This leads
to a scaling prediction of $s^{-8}$ for fixed angle meson-proton
scattering.

\vskip\baselineskip
\noindent
{\bf Acknowledgements:} This work has been partially funded through the
European TMR Contract No.~FMRX-CT96-0008. T.F. is supported by
Deutsche Forschungsgemeinschaft. P.K. thanks DESY Zeuthen for support
and the hospitality extended to him.

\end{fmffile}
\end{document}